# From Technical Debt to Cognitive and Intent Debt:
*Rethinking Software Health in the Age of AI*

*Margaret-Anne Storey*
*University of Victoria, Canada*
*mstorey@uvic.ca*

Generative AI is dramatically accelerating the velocity of software development, enabling small teams to ship features at a pace that would have seemed implausible just a few years ago [Peng et al. 2023]. I saw this firsthand in an entrepreneurship course I taught recently. Student teams were building software products over the semester, moving quickly to ship features and meet milestones. By week eight, one team hit a wall. Simple changes were breaking things in unexpected places, and progress had stalled. When I met with them, they initially blamed technical debt: messy code, hurried implementations, architectural shortcuts. But as we dug deeper, a different problem emerged. No one on the team could explain why certain design decisions had been made, or how different parts of the system were supposed to work together. The code might have been messy, but the deeper issue was that the team's shared understanding, the theory of the system [Naur 1985], had quietly fragmented. They had also failed to write down or communicate the rationale behind decisions. They had accumulated cognitive and intent debt faster than technical debt, and it had paralyzed them.

This is not an isolated story [Willison 2026]. Generative AI does not remove the challenges of software engineering; it redistributes them. In this article, I propose a **triple debt model for reasoning about software health**, built around three interacting debt types: **technical debt** refers to problems in the code layer, **cognitive debt** refers to erosion of shared understanding across a team over time, and **intent debt** refers to a lack of externalized goals, constraints, and rationale that both humans and AI systems need to work safely and efficiently with the codebase. Technical debt makes systems harder to change. Cognitive debt makes systems harder to understand. Intent debt makes it difficult to know what the system is actually for.

## The Hidden Cost of AI-Generated Code

For years, software engineers have worried about **technical debt** [Cunningham 1993], the long-term cost of messy code that accumulates and compounds when teams prioritize speed over quality. But generative AI may be quietly shifting where the real risk lies. Today, AI systems can generate code faster than developers can read or understand it, and as those models improve, the systems show growing promise for reducing debt through automated refactoring, test generation, and automated code review [Hou et al. 2024]. That is, generative AI may reduce technical debt while simultaneously accelerating the accumulation of cognitive and intent debt. Organizational expectations for rapid productivity gains, without corresponding investment in learning support, can create a paradoxical effect where developers lack the time to develop the understanding that would actually save them time [Miller et al. 2026].

The code may work. It may even be well-architected. But the team may not adequately understand *how* it works or remember *why* it was built the way it was. Over time, the shared understanding that makes a software system safe to change quietly erodes. This gradual loss of understanding increases **cognitive debt,** while the loss of captured rationale leads to **intent debt.** In AI-assisted development, cognitive and intent debt may quietly become the risks that matter most.

## What is a Software System, Really?

When developers think about a software system, they typically think first about code: the files, functions, and architecture that implement its features. But a software system exists across three distinct layers:

→ **Goals and intent:** the requirements, constraints, and objectives the system is meant to fulfill, as held by stakeholders and captured in specifications, tests, and documentation.

→ **Code and structure:** the representation of a system and the implementation of its intent: the source code, architecture, dependencies, and deployment infrastructure that make the system executable.

→ **Shared understanding:** the dynamic mental models that developers, architects, product managers, and other stakeholders hold about how the system works and how it can be reasoned about, what Naur called the "theory of the system" [Naur 1985]. Importantly, no single individual needs to understand all aspects of these layers; what matters is that sufficient shared understanding exists across the team to support safe change and coordination.

Technical debt has long been used to describe problems in the second layer. But the health of a software system depends on the alignment and quality of all three layers. When intent is unclear, systems drift from their intended purpose. When shared understanding is inadequate, teams cannot reason about change safely and are even reluctant to make changes. Tests may pass, but the product may have the wrong behavior. These failures are not only about technical debt in the code, although technical debt remains a concern.

## Technical Debt: The Well-Known Layer

Technical debt arises when developers make deliberate or inadvertent trade-offs that prioritize short-term delivery over long-term code quality. They take shortcuts that lead to messy or smelly code and make (or avoid) architectural decisions that constrain future evolution [Kruchten 2012]. The argument is that like financial debt, technical debt accrues interest: the longer it goes unaddressed, the more costly it becomes to change the system.

Technical debt, although problematic, is perhaps the easiest to manage of the three types of debt discussed in this article, in part because of its visibility. And the software engineering community has developed a rich set of accepted and validated practices to manage technical debt [Beck 1999, Fowler 2003]: **test-driven development, refactoring and code review,** among others. Generative AI is increasingly contributing to this work to reduce technical debt, providing automation for refactoring, identification of code or architectural smells and test generation. This optimism aside, even a codebase with low technical debt can be deeply problematic if the team does not understand it, or if it no longer reflects what the system was meant to do.

## Cognitive Debt: The Invisible Layer

*Cognitive debt is a team-level, project-level property reflecting the erosion of shared understanding across a software system over time.* This erosion manifests as increasingly inadequate mental models that developers rely on to reason about the system and change it safely and confidently. While individual developers experience this erosion as confusion, loss of control, or reduced confidence [Starr & Storey 2026], the debt itself resides in the accumulation of gaps in shared understanding and awareness across the team.

The term has also been used to describe measurable reductions in individual neural engagement during AI-assisted tasks [Kosmyna et al. 2024]. Related work describes 'comprehension debt' as the growing gap between code developers can produce with AI and what they genuinely understand [Alakmeh et al. 2026]. Our use of the term, however, focuses on the team-level and longitudinal dimension of software development: the *accumulated* erosion of shared understanding of a software system over time.

Developers have always struggled to understand not just legacy systems they did not build [von Mayrhauser & Vans 1995], but also software they built themselves using libraries they did not fully grasp, or Stack Overflow solutions they copied without comprehending. Classic case studies also observed that software systems depend on distributed human understanding, no one developer understands all of a system [Curtis et al. 1988]. Software development has never required any one person to understand an entire system; rather, it depends on sufficient, shared understanding across the team to enable safe change, coordination, and accountability. Cognitive debt is not new; developers have long worked with incomplete and distributed understanding of complex systems. What is new is the rate at which this gap can accumulate, and the difficulty of detecting it in AI-assisted development.

Generative AI further changes the relative importance of cognitive debt with respect to technical debt in a fundamental way. When a developer writes code from scratch, even messy code, the friction and effort mean they build at least a partial mental model along the way. They strive to understand what the code is trying to do, even if the implementation is imperfect. When an AI generates that same code, the developer may accept it without building the same level of understanding. At scale, across a team, and over time, this creates an accumulation of not knowing across the team. The code works, but the understanding and mental models of how the system behaves and how to reason about it are missing or flawed [Starr & Storey 2026].

## *Cognitive Surrender Leads to Cognitive Debt*

Reading and reasoning about unfamiliar code is among the most cognitively demanding activities developers undertake [Hermans 2021], which is partly why understanding erosion is so consequential. The psychological mechanism behind this is what Shaw and Nave [2026] refer to as **cognitive surrender**: adopting AI outputs with minimal scrutiny, bypassing both intuition and deliberate reasoning (drawing on Kahneman's [2011] distinction between fast, intuitive System 1 thinking and slower, more deliberate System 2 reasoning). This is importantly different from cognitive offloading, the strategic delegation of a discrete task to a tool (using a linter, a type checker, a spell-checker). Cognitive offloading is a rational productivity choice. Cognitive surrender can also be intentional but it may also lead to the loss of critical engagement. Even when surrender is intentional, the resulting debt accumulates invisibly. The team doesn't realize what understanding they're losing until it's gone. Notably, Shaw and Nave find that cognitive surrender also inflates confidence even when the AI is wrong, which helps explain why cognitive debt remains invisible until it is too late: the team feels they understand the system better than they do.

Cognitive debt, like technical debt, may also accumulate deliberately based on a conscious economic trade-off between understanding and speed of delivery. In my entrepreneurship course, students were rightly focused on getting rapid feedback on their ideas and cared less about technical debt. In so doing, they also accumulated cognitive debt, failing to maintain a shared understanding of where they were headed. In larger organizations, keeping track of an acceptable amount of cognitive debt can be even more challenging, because of its mostly invisible nature and scale. Effective problem-solving requires continuous interaction between a developer's model of the problem and their model of the system, each informing and correcting the other. When AI handles solution-building, this feedback loop is severed, leaving developers with an increasingly static understanding of the problem itself [Tornhill 2025], while also losing track of who knows what across the team.

## *Diagnosing Cognitive Debt*

Cognitive debt is harder to see and measure than technical debt because it is distributed across people and not directly visible in the codebase. But it leaves some visible traces and signals:

- **Resistance to change**: low confidence in understanding the system may make developers across the team reluctant to modify it [Starr & Storey 2026].

- **Unexpected results:** a reliable signal of cognitive debt is when a team member makes a change expecting one set of observable outcomes, but sees something else entirely [Starr & Storey 2026].

- **Slow or unpredictable onboarding**: new team members cannot get up to speed despite documentation, because the documentation describes what the code does, not why, and others on their team also struggle to help newcomers onboard.

- **Loss of transactive memory** [Wegner 1987, Herbsleb 2007]: team members lose track of who knows what and how the system is understood, increasingly relying on individual interactions with AI rather than shared practices.

- **Low bus factor**: only one person or very few on the team truly understands the system and other team members are concerned what may happen if they leave [Tornhill 2015].

### *Practices for Mitigating Cognitive Debt*

Some of the practices that work for technical debt are also useful here: **human code review**, for example, is valuable not only for catching defects but for spreading understanding across the team. **Pair programming** serves a similar function. But cognitive debt also calls for specific and new team understanding practices:

- **system walkthroughs** where developers explain code they did not write for the purpose of knowledge sharing and theory building [Naur 1985] rather than documentation or code refactoring,

- **retrospectives and post mortems** when things break or challenges emerge, to help collectively rebuild mental models that may have frayed,

- **deliberate communication during implementation, onboarding and offboarding** that surface gaps in shared understanding through grounding interactions [Clark 1991], and

- **reimplementation to repair cognitive debt**, as the cost of code generation is now relatively "low", teams can instruct an agent to reimplement a feature using different tests or new design elements that may help rebuild understanding.

While considering these practices, a common theme emerges: the practices that most effectively reduce cognitive debt are those that make implicit knowledge explicit, not as an afterthought, but as part of the development process itself. Some practitioners also propose using **AI** to reduce the cost of these practices and aid in understanding, but they should not fully replace the human friction involved in understanding. The need to externalize motivates the recognition of a third layer of debt, discussed next.

## Intent Debt: The Forgotten Layer

There is a third layer of debt that has received less attention, but which becomes increasingly critical as AI systems take on larger roles in software development: *intent debt refers to the absence or erosion of explicit rationale, goals, and constraints that guide how a system evolves.* Intent debt is not just about missing context and documentation. It accumulates when the goals, constraints, and specifications that should guide a system are unclear, poorly articulated, or not captured in any artifact that humans, and increasingly, AI agents can consult. What some practitioners are beginning to call "context debt", the missing information AI agents need to work effectively,  often includes symptoms of intent debt.  This growing urgency of tools to capture intent is what Marshall McLuhan described as technological "retrieval" [McLuhan & McLuhan 1988]: the introduction of new tools often revives practices that earlier technologies made less necessary or even obsolete. As generative AI systems generate more code, practices that capture intent, such as specifications, tests, domain knowledge, and design rationale, may become critical again.

Where technical debt lives in code and cognitive debt lives in people, intent debt lives in incomplete or missing non-code artifacts. Artifacts such as requirements documents, architectural decision records, implementation plans, tests, and specifications are the externalized memory of what a system is supposed to do. When this information is absent, incomplete, or out of date, cognitive debt cannot be repaired, and the system may gradually drift from its intended purpose and neither the team of developers nor the AI agents assisting them will have a reliable source of guidance.

Intent debt is also a familiar challenge in software engineering, especially for larger teams. Software projects have always struggled with requirements that drift; specifications and decisions that are never written down; and goals and constraints that exist only in the minds of a few key stakeholders. Intent is best captured at the moment key decisions are made, as recovering it later can be difficult and sometimes impossible (unlike technical debt, which can be addressed later). The arrival of AI agents as active participants in software development makes addressing intent debt more urgent than before, both for developers who may have inadequate understanding and for AI agents that may need to refactor, extend, or test a system.  AI agents and human developers may need to understand what the system is for, not just what it currently does. Without this information, human developers may find it impossible to build sufficient shared understanding, and agents may optimize for the wrong objectives. Each generation of AI-assisted development not only carries the debt forward but compounds it.

### *Diagnosing Intent Debt*

Intent debt can be recognized by several patterns:

- **Behavior drift:** the system's behavior diverges from what stakeholders believe it should do, discovered only during early testing or worse during customer incidents.

- **AI agents struggle to make changes:** Just as AI agents can struggle with poor quality code [Borg et al. 2026], AI agents may require extensive clarification, may produce solutions that are technically correct but miss the point, or use more tokens and time than expected because of a lack of context.

- **Loss of articulated constraints:** non-functional requirements such as performance budgets, privacy constraints, accessibility requirements, are known to a few people and are gradually forgotten.

### *Practices for Creating and Maintaining Intent Artifacts*

Before reliance on generative AI, source code often captured human intent through naming and design. But when code is generated by AI, intent may need to be deliberately articulated and captured through ***intent-first workflows.*** Reducing intent debt requires investing in "living" artifacts that externalize goals, plans, constraints, and reasoning:

- **Executable intent**. BDD specifications and tests designed to capture purpose rather than just verify behavior [Wynne 2017]. These are intent artifacts that can be executed and when they fail, they may show that the system has drifted from its intent.

- **Decision and rationale records.** Architectural Decision Records (ADRs) capture what was decided, why, and what was not done [Nygard 2011]. Domain-Driven Design (DDD) offers a complementary approach [Evans 2003]: its practices of ubiquitous language and collaborative domain modelling, make domain intent explicit before it is encoded in code.
- **Context artifacts for AI-assisted development.** Context engineering (skills, agent instructions, and playbooks) [Böckeler 2026] and AI-assisted intent capture from meetings and conversations [Ulloa et al. 2026] are emerging practices producing artifacts to be used by both humans and agents.

Despite their promise, these approaches cannot replace the hard human work of making decisions and exercising creative judgment about what software solution best serves users [Petre and Shaw 2025].

## Three Layers of Software System Health

Together, technical, cognitive, and intent debt  point to a framework for thinking about software system health that goes beyond code quality and changeability:

**Technical debt** lives in code. It accumulates when implementation decisions compromise future changeability. It limits how systems can change.

**Cognitive debt** lives in people. It accumulates when shared understanding of the system erodes faster than it is replenished. It limits how teams can reason about change.

**Intent debt** lives in artifacts. It accumulates when the goals and constraints that should guide the system are poorly captured or maintained. It limits whether the system continues to reflect what we meant to build and it limits how humans and AI agents can continue to evolve the system effectively.

The three types of debt interact and reinforce each other. The distributed cognition theory describes how cognitive processes are properties of systems of people, artifacts, and their interactions [Hutchins 1995]. Intent debt can cause cognitive debt: when the purpose of a system is not well documented, new and returning team members cannot form accurate mental models of it. Conversely, developers who lack understanding of the system intent will not be able to externalize specifications and record decisions. Cognitive debt can cause technical debt: when developers do not understand a system, they are more likely to make poor implementation decisions. And technical debt can amplify cognitive debt: messy or poorly designed code is harder to reason about, diminishing understanding. The reinforcing dynamics among technical, cognitive, and intent debt are notable and each has the potential to mitigate or erode the other (see Figure 1).

Managing software system health therefore requires proactive attention to all three layers, not just the one that is easiest to measure.

## How AI Is Shifting the Balance

As mentioned, the three types of debt are not new, but generative AI is changing their relative importance and the rate at which they accumulate. For managing technical debt, AI shows great potential. Automated refactoring, AI-assisted code review, and AI-generated test suites can reduce technical debt accumulation and make it easier to improve legacy systems. If current trends continue, some believe AI may increasingly manage the code layer on behalf of developers, reducing the human cost of maintaining code quality.

AI can be used as a partial solution for mitigating cognitive and intent debt, but it may also be a risk multiplier if the humans involved surrender their cognition and do not proactively capture their intent. AI generates code faster than teams can build the understanding needed to safely change it. At least today, it accepts underspecified prompts, fills in gaps, and produces plausible-looking results that may miss the intent or system needs entirely. And as AI takes on more of the implementation and documentation work, the feedback loop and friction that traditionally forced developers to understand the code and think about their intent are weakened.

## Implications for Practice

What can practitioners do? The framework for system health presented in this paper suggests several practical priorities for software teams working with AI-assisted development:

→ **Treat understanding as a deliverable.** Just as working code is a product of software development, shared understanding should be treated as a first-class deliverable, something teams invest in explicitly rather than something that happens as a side effect of writing code. This means allocating time for understanding practices such as walkthroughs, retrospectives, and knowledge transfer during onboarding and offboarding, while also investing in tools and software design principles that help rebuild understanding.

→ **Intent-first workflows.** Capture intent early when using AI to assist or automate development. ADRs, well-written specifications, domain modelling, decision rationales, plans, and clear user acceptance criteria are the raw material that grounds human understanding and that AI agents will require to do useful work.

→ **Resist the automation of understanding.** There is a temptation to use AI to generate documentation as well as code, producing explanations of what the system does without the team ever building genuine understanding. This substitutes the appearance of understanding for the real thing and makes cognitive debt harder to detect. Teams should be cautious about any practice that produces the artifacts of understanding without hard cognitive work to build important mental models. Looking ahead, the core developer's skill may not be authoring code, but maintaining correct understanding of what the system does and why, and how it can evolve [Hicks 2024].

→ **Monitor the three layers in tandem.** Technical debt has a rich ecosystem of tools for measurement and monitoring. Similar attention should be paid to cognitive and intent debt: perhaps through onboarding time tracking, knowledge concentration metrics, requirements coverage analysis, and regular audits of the gap between documented intent and actual behavior. At the very least, a team may wish to reflect on which practices they use to minimize debt across the three interacting dimensions we mention in this article. A focus on only some practices may lead to trade-offs they do not realize (see Figure 1).

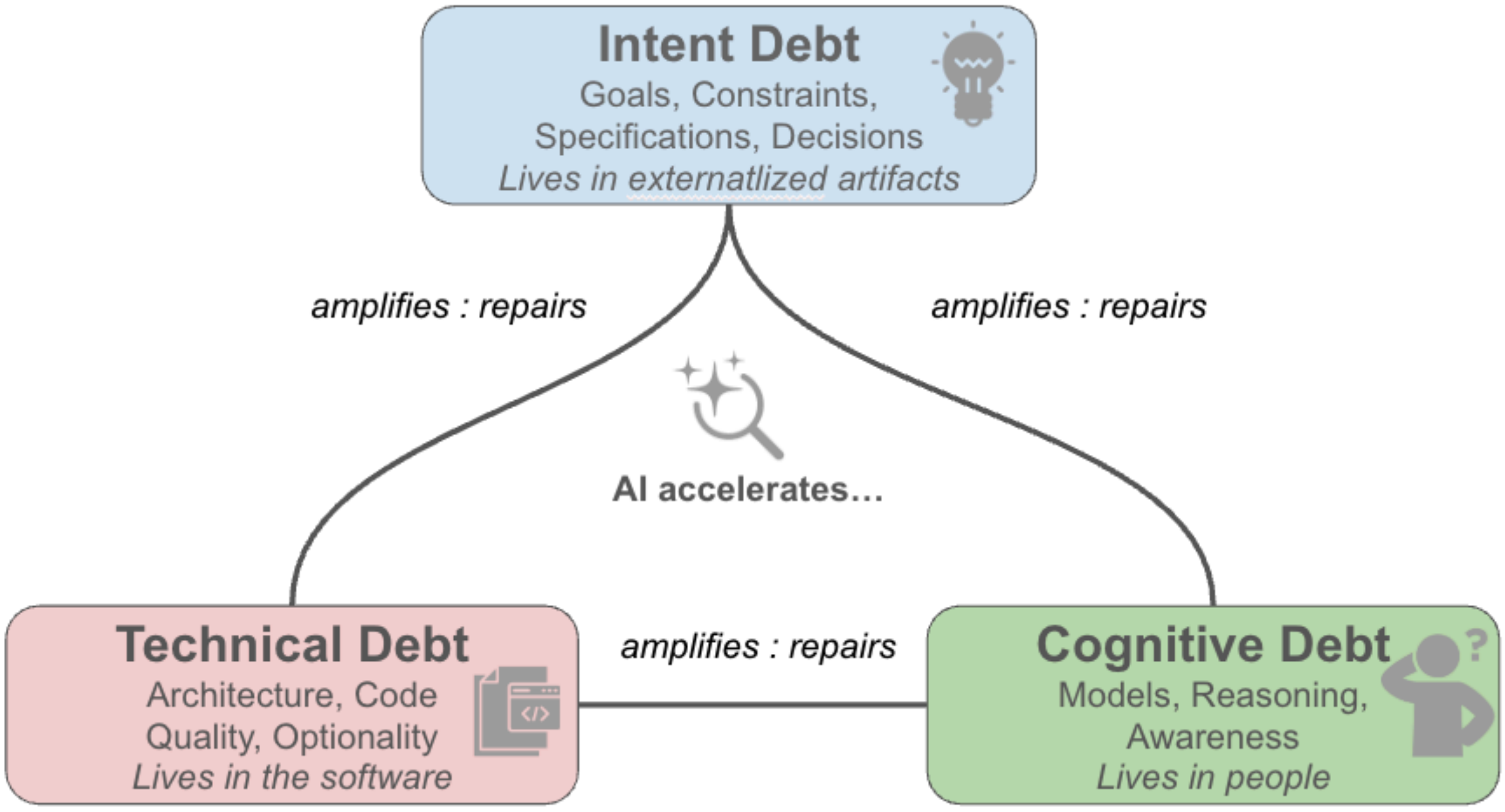

*Fig. 1: Software systems depend on: intent captured in artifacts, behavior embodied in code, and understanding distributed across the team and its artifacts. When these layers erode or fall out of alignment, three types of debt accumulate.*

## Points for Debate

### *Should we document "intent"?*

Not all practitioners agree that capturing intent is worth the effort, arguing that it doesn't matter how we got here; we just need to focus on what changes are needed to go forward and what we need to know can be (re)constructed. Others argue capturing intent is important, and generative AI may help. The debate about whether and how intent should be documented is far from settled.

### *Can AI help make implicit knowledge explicit?*

Some practitioners have proposed that AI can help surface implicit knowledge, reducing intent debt proactively and retroactively. Others counter that the main value in creating documentation is the human understanding gained. Whether and how AI should be used to autogenerate such documentation remains an open question.

### *Debt as a risk, or debt as a strategy?*

This paper frames cognitive debt as a risk to be managed, while acknowledging that understanding may not need to be about details of the implementation layer and may be distributed among team members and artifacts. Others argue this understanding is often not needed at all, similarly to how managers have always had to trust team members. With generative AI automating more development work, knowing how much debt is acceptable in a particular organization or project remains uncertain.

## Conclusion

Generative AI promises extraordinary gains in the speed and ease with which software can be created. But as with many technological shifts, the benefits may carry a hidden reversal [McLuhan & McLuhan 1988] from technical debt to cognitive and intent debt, with a need for tools and processes to support paying that down. The teams and organizations that navigate the AI-assisted development era most successfully will be those that invest in understanding and intent as deliberately as they have invested in code quality. Maintaining healthy software systems in the age of AI may depend on preserving alignment among intent, code, and understanding, and software teams should manage understanding with the same care and urgency as they manage code.

### Acknowledgments

I am grateful to colleagues who provided valuable feedback and ideas on reviewing earlier drafts of this work: Adam Tornhill, Kent Beck, Daniel German, Dave Thomas, Marian Petre, Markus Borg, Mary Shaw, Roy Weil, Max Alexander-Kanat, Keith Mann and Ciara Storey. I am grateful to Arty Starr for discussions that helped sharpen the definitions in this paper and for work that influenced my thinking about how understanding breaks down and is rebuilt in practice. I also thank colleagues who attended the Future of Software Engineering Retreat organized by Thoughtworks in Feb 2026 for their insights and feedback.